# Characterization of electronic structure of periodically strained graphene

*Marjan Aslani[1,*], C. Michael Garner[1,a)], Suhas Kumar[1,2,*], Dennis Nordlund[3], Piero Pianetta[1,3], and Yoshio Nishi[1]*

[1]*Department of Electrical Engineering, Stanford University, Stanford, California, 94305, USA*
[2]*Hewlett-Packard Laboratories, 1501 Page Mill Road, Palo Alto, California, 94304, USA*
[3]*Stanford Synchrotron Radiation Lightsource, SLAC National Accelerator Laboratory, 2575 Sand Hill Road, Menlo Park, California, 94025, USA*
*Equal contribution a) Corresponding author; email: mcgarner@stanford.edu

## Abstract

*We induced periodic biaxial tensile strain in polycrystalline graphene by wrapping it over a substrate with repeating pillar-like structures with a periodicity of 600 nm. Using Raman spectroscopy, we determined to have introduced biaxial strains in graphene in the range of 0.4% to 0.7%. Its band structure was characterized using photoemission from valance bands, shifts in the secondary electron emission, and x-ray absorption from the carbon 1s levels to the unoccupied graphene conduction bands. It was observed that relative to unstrained graphene, strained graphene had a higher work function and higher density of states in the valence and conduction bands. We measured the conductivity of the strained and unstrained graphene in response to a gate voltage and correlated the changes in their behavior to the changes in the electronic structure. From these sets of data, we propose a simple band diagram representing graphene with periodic biaxial strain.*

Graphene has been considered a promising material for advanced electronics with an electron mobility as high as 30,000 $cm^2V^{-1}s^{-1}$ at room temperature,[1,2] while its zero-bandgap property imposes a need for engineering and controlling of its electronic properties for device applications.[3,4] For graphene to be used as a semiconductor device with low off state leakage, it needs to have a bandgap and a high mobility, which is necessary for CMOS-like devices.[5,6] Strain engineering is currently used to enhance mobility in most silicon integrated circuits and is a common method to modify bandstructure.[7] Theoretical predictions claim that a considerable strain of ~20% can open a small bandgap.[8,9] Such predictions also claim that, for opening of a bandgap, a uniaxial strain along with shear strain,[9] or periodic strain[10] is required, while uniaxial strain alone produces no bandgap.[8,11] Here we chose to induce strain through a periodic deformation using a patterned substrate.

We chose $SiO_2$ as the substrate because its valence band is ~4 eV below the Fermi level, hence allowing for characterization of changes to the valence band of graphene using photoemission. The substrate was a grid array of $SiO_2$ nanopillars of 90-105 nm height, 200 nm wide, with a pitch of 600 nm, all placed on a 200 nm thick uniform $SiO_2$ film on a highly doped silicon wafer, as shown in Figure 1a. Commercially manufactured single layer graphene covered on top with PMMA was then placed with the graphene face down on the array of $SiO_2$ nanopillars. The PMMA was then driven off from the graphene with a forming gas anneal at 400 $^0$C. During this annealing process, the graphene was stretched over the $SiO_2$ surface and adhered to the $SiO_2$ in the "valleys" presumably by van der Waal's forces, as shown in Figure 1b (hence the stack sequence: Si/uniform $SiO_2$/pillars of $SiO_2$/graphene). Details on the fabrication process are described in the supplemental material.[12]

After the transfer of graphene onto flat $SiO_2$ (unstrained) and pillars of $SiO_2$ (strained), Raman spectra were collected using a beam size of ~1 µm at random positions over an area of 1 $mm^2$ and averaged to detect the presence of strain.[12]

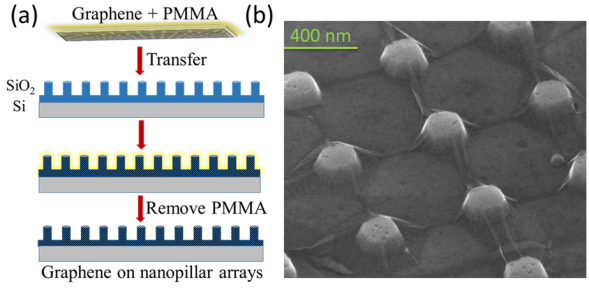

Figure 1: (a) Schematic depiction of graphene transfer process onto a nanopillar array. (b) Scanning electron micrograph of the 90 nm tall pillar structures with transferred graphene.

Graphene over pillars of height 90 nm and those of height 105 nm, showed shifts in the averaged Raman 2D peak of -23 cm$^{-1}$ and -35 cm$^{-1}$, in the D peak of -12 cm$^{-1}$ and -22 cm$^{-1}$, and in the G peak of -12 cm$^{-1}$ and -18 cm$^{-1}$, respectively, relative to graphene on flat SiO$_2$ (Figure 2a). These peak shifts correspond to approximately 0.4-0.7% of biaxial strain in each of the two lateral dimensions and equivalently 0.88-1.6% of uniaxial strain in either of the lateral dimension.[12-16] The deformation profile of graphene on nanopillars seen in Figure 1b suggests biaxial strain. For the rest of the measurements, strain was induced using pillars of height 90 nm.

X-ray photoemission spectroscopy was used to probe the valence bands of strained graphene. Photoemission from the samples is a composite of that from valence bands of both graphene and SiO$_2$. Since the valence band of SiO$_2$ is ~4 eV below the Fermi level, photoemission 0-4 eV below the Fermi level will be predominantly from graphene. Spectra of strained and unstrained graphene were measured under identical conditions and normalized in intensity to a common SiO$_2$ peak. Gold was evaporated around the area of interest on the sample and a potential of 25 V - 60 V was applied between the sample and the electron analyzer to sweep photo electrons from the textured sample surface to the electron analyzer. Graphene on flat SiO$_2$ had a photoemission peak centered at ~3 eV, (Figure 2b) associated with the π state of graphene. Strained graphene had photoemission peaks centered at the same energy, but had a significantly higher (by ~60%) magnitude than that on flat SiO$_2$.

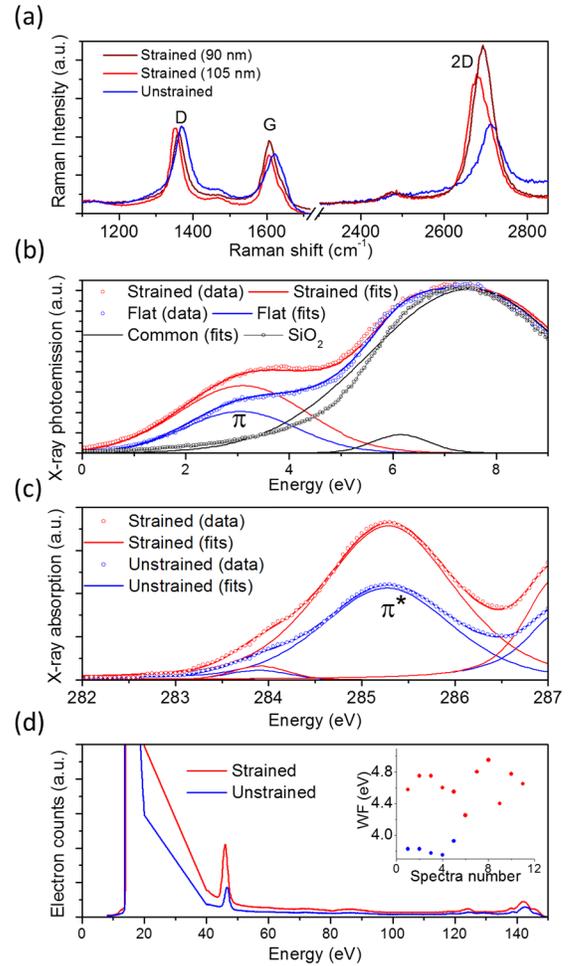

Figure 2: (a) Raman spectra of unstrained and strained graphene (on 90 nm and 105 nm tall pillars). (b) Photoemission from the valence band of unstrained and strained graphene, measured using a photon energy of 140 eV. The π band is identified with an annotation. Photoemission from SiO$_2$ with no graphene is also shown. The SiO$_2$ had undergone identical processing as the strained and unstrained graphene had. (c) X-ray absorption spectra from graphene 1s level to its π* band from unstrained and strained graphene. The π* band is identified with an annotation. (d) Photoemission spectra of unstrained and strained graphene on SiO$_2$ with an applied bias of ~10 V between the sample and the electron analyzer. Inset is the measured work function (WF) of unstrained and strained graphene using different spectra during different experimental measurements. The average work function of unstrained graphene was 3.8 eV and the average work function of strained graphene was 4.6 eV. Strained graphene in (b)-(d) refers to that on 90 nm tall pillars.

Increase in the π band density of states in strained graphene indicates the possibility that the band had broadened in the momentum space, consistent with theoretical claims.[10] This would imply a potential reduction in hole mobility with periodic strain, which is consistent with electrical characteristics shown later.

We employed near edge x-ray absorption fine structure (NEXAFS)[17] in x-ray absorption spectroscopy to study the conduction band properties of strained graphene, where we used the transitions from the C-1s to unoccupied antibonding orbitals of graphene. Spectra of strained and unstrained graphene were measured under identical conditions and normalized to a common background. In the NEXAFS of strained and unstrained graphene (Figure 2c), the peaks centered at 285-286 eV correspond to the π* states of graphene, or the lowest conduction band. The π* band in strained graphene appeared to have a significantly higher density of states (by ~60%) relative to that in unstrained graphene. Similar to the case of the conduction band, this increase in π* band density of states in strained graphene indicates the possibility that the band had broadened in the momentum space followed by a reduction in electron mobility. To determine the work function of strained and unstrained graphene, the kinetic energy difference between the Fermi level (103.3 eV above the $SiO_2$ Si-2p core level) and the secondary electron cut off was subtracted from the photon energy (Figure 2d). The resulting work function of strained graphene (4.6 eV) is ~0.8 eV higher than that of unstrained graphene (3.8 eV) (Figure 2d), consistent with a prior theoretical prediction.[18]

After the physical characterization described above, we measured the conductivity of strained and unstrained graphene in response to a gate voltage. For this, the $SiO_2$ (either patterned or flat) sandwiched between graphene and the highly doped substrate was used as the gate insulator and, the Si substrate was used as the gate electrode. Source-drain contacts of titanium and gold were lithographically defined. The sample was annealed in vacuum at 300 °C to eliminate possible trapped moisture beneath graphene that might harbor ions and affect the nature of the electrical behavior. The channel consisted of graphene stretched over several nanopillars (Figure 3a).

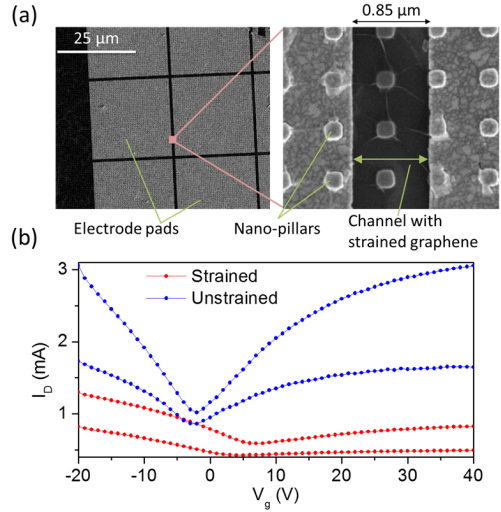

Figure 3: (a) Scanning electron micrograph of the fabricated field effect transistor structure used to measure the characteristics shown in (b). (b) Drain current ($I_D$) – gate voltage ($V_g$) characteristics of strained and unstrained graphene. Curves of same color are plots representing different devices that well represent the variability in $I_D$ –$V_g$ characteristics.

The gate voltage versus drain current curves displayed in Figure 3b shows that unstrained graphene is ambipolar in nature and is slightly n-type because the minimum drain current occurs at a negative gate voltage. In strained graphene, the ambiopolar behavior appeared to have been suppressed (also see Ref. 12) and the minimum drain current occurred at a positive gate voltage. The magnitudes of the currents from the two measurements themselves are not comparable because of differences in the capacitances and resistances associated with the two different structures in the strained and unstrained case.

The increase in the gate voltage required to achieve minimum drain current is consistent with and is probably accounted for by the increase in work function in strained graphene as suggested by spectroscopic data. The decrease in the transconductance by a factor of 2-3 (slope of drain current with respect to gate voltage) in the strained case, relative to the unstrained case, is an indication of decreased mobility and by implication, broadening of the bands. In fact, in the case of strained graphene, the gate oxide is half as thick as in the case of unstrained graphene, which means the transconductance is enhanced by a factor of 2 in strained graphene. Despite this, the decrease in transconductance in strained graphene indicates a decrease in mobility by a factor of 4-6. Since the centroids of the valence and conduction bands, as seen in the photoemission and x-ray absorption data, did not shift in energy upon application of strain, it is unlikely that the valence and conduction bands shifted in energy. In addition to the decreased mobility and an increase in the gate voltage required to achieve minimum drain current, there is also suppression of ambipolar behavior in the strained graphene at positive gate voltages. This means that, apart from the broadening of the bands and increase in the work function, there is an inability of the current carriers to access the conduction band at highly positive gate voltages. This suggests an increase in the effective bandgap, where there are no sufficient states to allow for a significant current flow. There are theoretical predictions of opening of a bandgap, broadening of bands in the momentum space and rounding of the sharp ends of the valence and conduction bands upon periodic strain, consistent with those observed here.[8-10] We incorporated the above aspects into a simple band diagram depicting the changes in strained graphene relative to unstrained graphene, displayed in Figure 4. This study also highlights that strained graphene may require a contact metal with a different work function than that used for unstrained graphene in order to form low resistance ohmic contacts. For instance, a metal with a work function of 0.8 eV more than Ti (WF = 4.3 eV) which was used in the electrical characterization here. Possible candidates include Cu<100> (WF=5.1 eV), Pt<331> (WF = 5.12 eV), polycrystalline Pd (WF = 5.22 eV), etc.

In conclusion, we applied periodic strain to graphene using a patterned array of $SiO_2$ nanopillars. Multiple spectroscopic techniques indicated that the biaxial strain was about 0.4% - 0.7% due to which the valence and conduction bands broadened while the work function increased by ~0.8 eV. Conductivity of strained graphene in response to a gate voltage was consistent with the spectroscopic analysis. We proposed a simple band diagram depicting the changes observed herein.

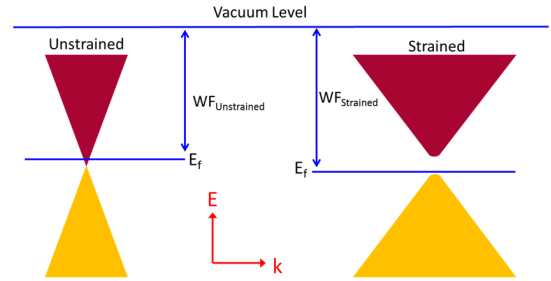

Figure 4: Qualitative schematic representation of strain-induced changes in the band structure of graphene. The two colors of the bands represent either sides of the Fermi level ($E_F$).

### Acknowledgements

The authors acknowledge the Semiconductor Research Corporation (SRC) for support under contract 2124.001. The authors also acknowledge Intel Corporation for support through the Center for Integrated Systems (CIS) at Stanford University. Use of the Stanford Synchrotron Radiation Lightsource, SLAC National Accelerator Laboratory, is supported by the U.S. Department of Energy, Office of Science, Office of Basic Energy Sciences under Contract No. DE-AC02-76SF00515. Work was performed in part at the Stanford Nanofabrication Facility which is supported by National Science Foundation through the NNIN under Grant ECS-9731293. Part of this work was performed at the Stanford Nano Shared Facilities (SNSF).